\documentclass[]{tJOT2e}

\newcommand{\eqNS}{
\partial _t \mathbf{u} + (\mathbf{u} \cdot \nabla) \mathbf{u} = -
\nabla P + \nu \triangle\mathbf{u} + \mathbf{f}
}

\newcommand{\Reyla}{$\mathcal{R}$\textit{e}$_\lambda$ }
\newcommand{\parfrac}[2]{\left(\frac{#1}{#2}\right)}
\newcommand{\eqdef}{\stackrel{\mathit{def}}{=}}

\newcommand{\gsim}{\stackrel{>}{\sim}}

\begin{document}
\doi{10.1080/14685240YYxxxxxxx}
 \issn{1468-5248}
 \jvol{00} \jnum{00} \jyear{2008}

\markboth{Taylor \& Francis and I.T. Consultant}{Journal of Turbulence}

\articletype{}

\title{Geometric properties of particle trajectories in
  turbulent flows}

\author{A. Scagliarini$^{\rm a}$$^{\ast}$\thanks{$^\ast$Corresponding
    author. Email: andrea.scagliarini@roma2.infn.it}\\{\em{Department of
    Physics, University of Rome ``Tor Vergata'',\\ 
    Via della Ricerca Scientifica 1, 00133 Rome, Italy}} }

\maketitle

\begin{abstract}
We study the statistics of curvature and torsion of Lagrangian
trajectories  from direct numerical simulations of homogeneous and isotropic turbulence (at \Reyla$\approx 280$) in order to extract informations on the geometry of small scale coherent structures in
turbulent flows.
We find that, as previously observed by Braun \textit{et al}
\cite{braun} and by Xu \textit{et al} \cite{xu}, the high curvature statistics is dominated
by large scale flow reversals where the velocity magnitude assumes very
low values. In order to focus on small-scales signatures, we introduce
a \textit{cutoff} on the velocity amplitude and we study
the  probability distribution of time--filtered curvature conditioned only on those 
events when the local velocity is not that small. 
In this way we are able to select small--scales turbulent  features, 
connected to vortex filaments. 
We show that the conditional probability density of time--filtered
curvature  is well
reproduced by a multifractal formalism. 
Finally, by studying  the joint statistics of curvature
and torsion we find further evidences that intense and persistent events are dominated by helical type trajectories. 
\bigskip

\bigskip

\end{abstract}

\section{Introduction}
The transport of particles in turbulent flows is an ubiquitous
phenomenon relevant in very different physical contexts as cloud
formation, pollutant dispersion, formation of ``planetesimals'' and
turbulent mixing in chemical reactions
\cite{pope,yeung,falko-fouxon,post}.  To decribe it, the optimal
framework results to be the Lagrangian one, where flow properties are
studied following the trajectories of individual fluid elements. Thus,
understanding the Lagrangian statistics of particles advected by a
turbulent velocity field $\mathbf{u}(\mathbf{x},t)$ is a fundamental
issue for both its theoretical implications (as, for instance, the
development of stochastic models for dispersion and mixing) and for
applications \cite{pope,sawford2001,arneodo_ictr}.  Lagrangian
description is also more suitable than the Eulerian one to reveal the
geometry and the statistical ``weight'' of coherent structures in
turbulent flows.  We cite here, for example, the preferential
concentration of heavy/light particles inside hyperbolic/elliptic
regions \cite{eaton,bif-bec,bec-bif}, the strong acceleration bursts
experienced by tracers when following streamlines around vortex
filaments \cite{biferale-trapp} or the regularising effects of inertia
in the transition from viscous scales to inertial range scales in the
behaviour of Lagrangian Structure Functions \cite{bif-bec}.  Studies
of Lagrangian turbulence mainly involve the analysis of the statistics
of velocity (or vorticity) and acceleration
\cite{yeung,laporta,pinton,ott-mann,tsinober,biferale-trapp} (for an
updated recent review see Toschi \& Bodenschatz \cite{toschi-bod}).
As suggested by Braun \textit{et al.} \cite{braun}, the
trajectory of a Lagrangian particle $\mathbf{x}=\mathbf{x}(t)$ can be
also seen as a set of parametric equations representing a spatial
curve in 3D. As such it can be described by means of tools borrowed
from differential geometry: in particular, it is well known that a
curve is uniquely determined by the knowledge at every point (or
equivalently, in our ``time--like'' parameterization, at every instant
of time) of two fundamental geometric parameters: the curvature
$\kappa$ and the torsion $\vartheta$. They are related to the velocity
($\mathbf{u}$) and the acceleration ($\mathbf{a}$) of the particle by:
\begin{equation} \label{eq:curv}
\kappa(t)=\frac{|\mathbf{u}\wedge\mathbf{a}|}{u^3}=\frac{a_\perp}{u^2}
\end{equation}
(where $a_\perp$ is the magnitude of the normal, or centripetal,
acceleration) and
\begin{equation} \label{eq:tors}
\vartheta(t)=\frac{\mathbf{u}\cdot(\mathbf{a}\wedge\mathbf{\dot{a}})}{\kappa^2u^6}.
\end{equation}
The torsion, unlike the curvature, has got a sign, which is related to
the local helicity \cite{moffatt,sposito}. Since we are not dealing with
helical flows (that is  we have a symmetric distribution of regions
with positive and negative helicity, hence the same is for torsion),
from now on we will refer to torsion discussing its absolute value
$\theta= |\vartheta|$.
Dimensionally $\kappa$ and $\theta$ are an inverse length, so we may
expect that they assume very high values in correspondence of the
small scale structures we are interested in, namely the vortex
filaments.  It has been shown
\cite{laporta,biferale-lagr,biferale-trapp,bif-bec} that such
filaments are of great relevance in the Lagrangian statistics of
turbulent velocity and acceleration for both tracers and inertial
particles (playing in the latter case a crucial role in the mechanism
responsible for the strong inhomogeneous particle spatial
distribution). The aim of this work is twofold.
First, we will further characterize the geometry of
these coherent structures by studying the statistics of extreme values
of curvature and torsion from Direct Numerical Simulations, 
extending the previous analysis made in \cite{braun} to higher
Reynolds numbers and supporting the numerical results with a
phenomenological description 
in terms of the multifractal theory. Second, thanks to the high
Lagrangian statistics, we will also present joint statistics of
torsion and curvature, 
highlighting the intimate geometrical structure of small scales intense vorticity in turbulent flows. 

The paper is organized as follows. In section \ref{sec:curv_stat} we
discuss the statistics of instantaneous and time--averaged (and
conditioned on velocity values) curvature, in the latter case deriving
some analytical results in the framework of the multifractal
formalism; in section \ref{sec:joint_stat} we analyze the joint
statistics of curvature and torsion in order to stress its connection
with the topology of small--scale coherent structures. Conclusions and
perspectives are left to the last section.

\section{Curvature statistics} \label{sec:curv_stat}
We analyzed data produced in a Direct Numerical Simulation of
 homogeneous and isotropic turbulence in 
 a cubic lattice of $1024^3$ grid points \cite{biferale-trapp}, corresponding to $\mbox{\Reyla}=284$, seeded
with about two millions Lagrangian particles, evolving according
to the dynamics
\begin{equation} \label{eq:trac.motion}
\frac{d\mathbf{x}(t)}{dt}=\mathbf{u}(\mathbf{x}(t),t),
\end{equation}
where $\mathbf{u}(\mathbf{x},t)$, the fluid velocity field, 
is a solution of the incompressible 
Navier--Stokes equations
\begin{equation} \label{eq:NS}
\eqNS 
\end{equation}
$$
\nabla\cdot\mathbf{u}=0
$$
which have been integrated on a triply periodic cubic box by means of
a fully dealiased pseudospectral code (energy was injected at an
average rate $\varepsilon$ by keeping the total energy constant in
each of the first two wavenumber shells \cite{chen}). The equations
(\ref{eq:trac.motion}) have been integrated linearly interpolating the
values of the fluid velocity at the lattice sites.  Let us stress that
the acceleration field we can measure is continuous (and
differentiable inside lattice mesh) because it is reconstructed from
the local values of pressure gradients, laplacian of velocity and the
external forcing along the trajectories. In such a way  we can calculate
$\mathbf{a}$ directly from the rhs of (\ref{eq:NS}), without
performing any differentiation of the particle velocity, thus
minimizing noise effects, which are
unavoidably present in experiments and, in general, in numerical
simulations as well.

\subsection{Instantaneous and ``time--filtered'' curvature} \label{subsec:curv_stat.1}
\begin{figure}
\begin{center}
\resizebox*{10cm}{!}{\includegraphics{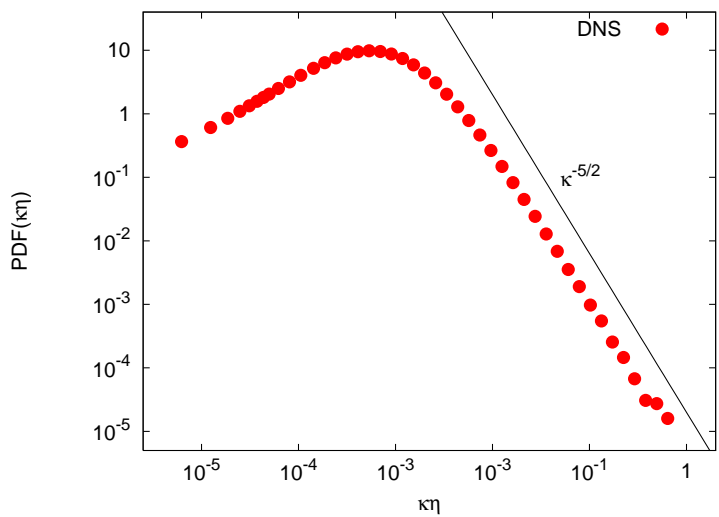}}
\caption{Log--log plot of the curvature PDF, normalized with the
  dissipative scale $\eta$. 
The solid line represents the $\kappa^{-5/2}$
 scaling of the PDF for high curvature values. 
Data, here and hereafter, come from a high resolution DNS at $1024^3$ collocation points, see text for details.}
\label{fig:pdfkslope}
\end{center}
\end{figure}

\begin{figure}
\begin{center}
\resizebox*{10cm}{!}{\includegraphics{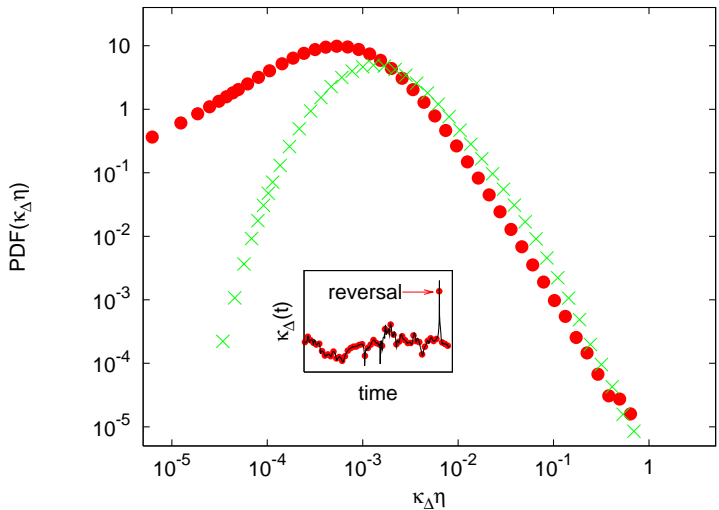}}
\caption{Log--log plot of PDFs of instantaneous ($\bullet$) and time averaged ($\times$) 
   curvature (over a window $\Delta=10\tau_{\eta}$). It must be noticed
   that only the tail at
   low $\kappa$ values has been effectively damped down by the filtering. \textit{Inset}:
 Time series of instantaneous (solid line) and time--filtered ($\bullet$) curvature relative to a trajectory in the 
  high $\kappa$ tail of $\mathcal{P}(\kappa_{(\Delta=10\tau_{\eta})})$: the arrow highlights
 the strong flow reversal event which is only very weakly damped by time averaging.}
\label{fig:pdfk.filt.nofilt}
\end{center}
\end{figure}

In figure (\ref{fig:pdfkslope}) we show the PDF of the instantaneous
curvature $\mathcal{P}(\kappa)$.
It is possible to notice the same power--law tail
$\mathcal{P}(\kappa)\sim\kappa^{-5/2}$, for large curvatures, observed
in the numerical simulations by Braun \emph{et al.} \cite{braun} and
in the experiments by Xu \emph{et al.}  \cite{xu}. In the latter work
it is argued that such a behaviour can be explained considering that
the very high curvature events are linked to large scale flow
reversals where the particle inverts its motion and its velocity goes
to zero, $u^2\sim0$.  Therefore, the PDF scaling for large curvatures
is mainly determined by the Gaussian statistics of the velocity and,
in this context, it can be reproduced analytically by simple
arguments \cite{xu}.
\\
In order to avoid contamination from large scale motion, in \cite{xu} 
 it was suggested to enhance the contribution of vortex
filaments by  studying the statistics of the curvature averaged over
some time intervals $\Delta$ \cite{bif-toschi}. The \emph{rationale}
at the basis of this operation consists in the observation that, while
the ``trapping'' events of particles into the filaments are
characterized by a certain degree of temporal persistency (due to the
already stated spatio--temporal coherence of such structures), the
flow reversals with very low velocity show very little
autocorrelation times. Indeed, as also remarked in \cite{xu},
the larger is the curvature the lesser is
the autocorrelation time (in table \ref{tab:ctimes} we report such 
times extracted from the autocorrelation function of curvature,
conditioned on its magnitude, together with the corresponding conditioning
threshold). 
\begin{table}
\begin{center}
\begin{tabular}{|l|l|}
\hline 
 $\log (\kappa^{\ast})$ & $ \tau_{corr} $ \\
\hline
  $1.5$ & $3.8\tau_\eta$ \\
\hline
  $2.5$ & $1.56\tau_\eta$ \\
\hline
  $3.5$ & $0.12\tau_\eta$ \\
\hline
\end{tabular}
\end{center}
\caption{Conditional autocorrelation times and corresponding
  conditioning thresholds. The autocorrelation times have been
  extracted from the conditional autocorrelation function of the
  logarithm of curvature, defined as 
$$
\mathcal{C}(\tau) = \frac{\langle
  \log(\kappa(t+\tau))\log(\kappa(t))|\log(\kappa(t))>\log(\kappa^{\ast}) 
\rangle - \langle\log(\kappa(t))|\log(\kappa(t))>\log(\kappa^{\ast})\rangle^2}{\langle
  \log(\kappa(t))^2|\log(\kappa(t))>\log(\kappa^{\ast})\rangle - \langle\log(\kappa(t))|\log(\kappa(t))>\log(\kappa^{\ast})\rangle^2} \sim e^{-\tau/\tau_{corr}}.  
$$
 At increasing the threshold (including in the computation
 only higher and higher curvature values) the autocorrelation
 decreases significantly.}
\label{tab:ctimes}
\end{table}
We have, thus, measured the PDFs of the
time-filtered curvature:
\begin{equation} \label{eq:kfilt}
\kappa_\Delta(t)\eqdef\frac{1}{\Delta}
\int_{t-\Delta/2}^{t+\Delta/2} \kappa(t^\prime)dt^\prime
\end{equation}
for $\Delta=\tau_\eta,5\tau_\eta,10\tau_\eta$, where $\tau_\eta$ is
the Kolmogorov time scale. However, even for the greatest time
``window'' ($10\tau_\eta$), we do not obtain the desired result. In
fact, as it is evident from figure (\ref{fig:pdfk.filt.nofilt})
where we report the PDFs of instantaneous and time--filtered (with
$\Delta=10\tau_\eta$) curvature, time--filtering is
effective mostly at small $\kappa$, where the PDF exhibits an
important lowering of the tail.
The explanation why filtering is not effective at high $\kappa$ values
can be understood with the help of a single example. In the inset of
figure (\ref{fig:pdfk.filt.nofilt}) we analyse the effect of filtering
around a very intense event in our statistics. We superimpose the
temporal signal of the curvature together with the filtered one for a
trajectory selected at random in the
$\mathcal{P}(\kappa_{(\Delta=10\tau_\eta)})$ tail: at the instant of
time at which the event of flow reversal occurs (indicated with an
arrow) the curvature becomes so large that it is weakly damped even
averaging over the longest time interval we used.

\subsection{Vortices identification and analytical results} \label{subsec:vortices} 
\begin{figure}
\begin{center}
\resizebox*{10cm}{!}{\includegraphics{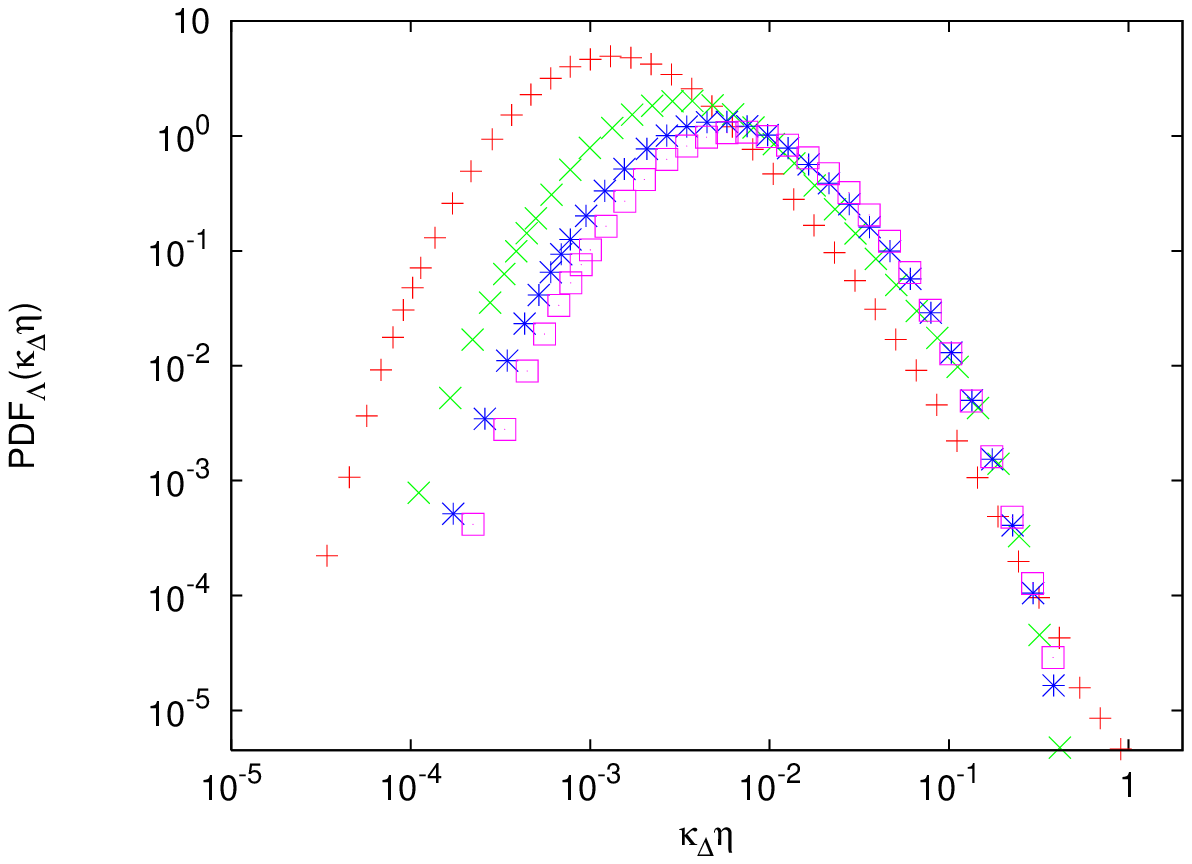}}
\caption{Log--log plot of conditional PDFs (of the time--averaged curvature),
  computed including only those events for which $u^2 > \Lambda$, for
  $\Lambda = 0.02u^2_{rms}$ ($\times$), $0.1u^2_{rms}$
  ($\ast$) and $0.2u^2_{rms}$
  ($\Box$), respectively. The introduction of the
  cutoff induces a change in the slope of the far right tail (we also
  plot the unconditioned PDF ($+$) for comparison),
  reaching an asymptotic shape for the right tail insensitive to the
  increasing of  $\Lambda$.}%
\label{fig:cutoffs}
\end{center}
\end{figure}

\begin{figure}
\begin{center}
\resizebox*{10cm}{!}{\includegraphics{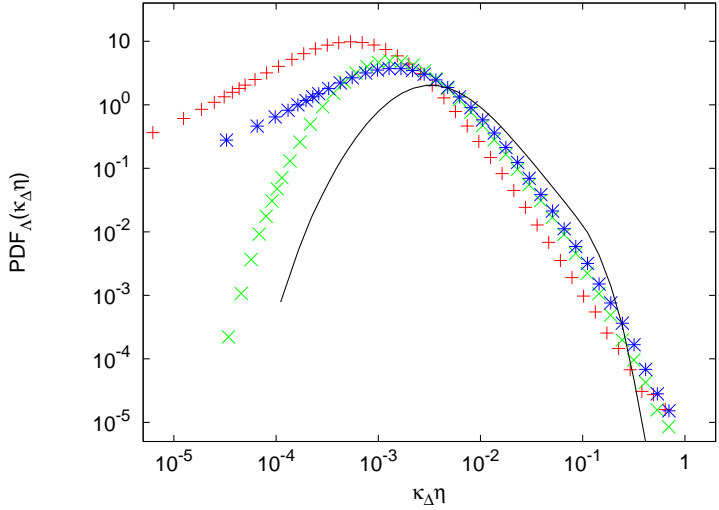}}
\caption{Log--log plot of the PDFs of instantaneous curvature without ($+$) and 
with ($\ast$) the cutoff on velocity values, and of the PDFs of
time--filtered curvature without ($\times$) and with (solid line) the cutoff. 
 It is important to notice that
 the change of the PDF slope for large curvatures is due to the
 combination of the two operations (averaging in time and constraining
 the velocity magnitude).}%
\label{fig:pdfks}
\end{center}
\end{figure}

\begin{figure}
\begin{center}
\resizebox*{10cm}{!}{\includegraphics{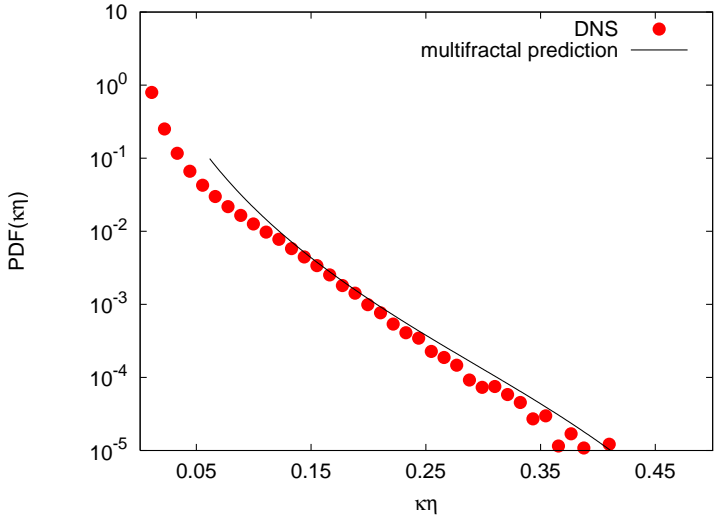}}
\caption{Log--lin plot of the PDF with cutoff on $u^2$ of the time filtered
   curvature ($\Delta=10\tau_\eta$) and the corresponding multifractal prediction.}%
\label{fig:pdfktail}
\end{center}
\end{figure}

In order to really focus only on
those high--curvature events given by large values of the numerator in
(\ref{eq:curv}) we propose to introduce also  a \emph{cutoff} ($\Lambda$) on
the values of the velocity, i.e.  we included in the computation of
the filtered PDF only those events for which
\begin{equation} \label{eq:cutoff}
u^2 > \Lambda.
\end{equation}
We have tried various values $\Lambda$,  see
fig. (\ref{fig:cutoffs}), and observed  that there is a
tendency toward a  PDF shape with a right tail insensitive to the increasing of the cutoff; keeping this in
mind we will fix in what follows $\Lambda=0.02 u^2_{rms}$.
In this way, we define a \emph{conditional} PDF:
\begin{equation}
\mathcal{P}_\Lambda(\kappa_\Delta)\eqdef
\mathcal{P}(\kappa_\Delta|u^2>\Lambda).
\end{equation}  
As one can observe from figure (\ref{fig:pdfks}), the application of this
constraint results in a dramatic change in the shape of the PDF for
high curvature values; in particular, the change of slope of 
$\mathcal{P}_\Lambda(\kappa_{(\Delta=10\tau_\eta)})$ appears
around the value
\begin{equation} \label{eq:threshold}
\tilde{\kappa}\eta \equiv \kappa_{(\Delta=10\tau_\eta)}\eta\approx 0.1
\end{equation} 
where $\eta$ is the Kolmogorov length.  
This new shape for the conditional PDF of the time--filtered curvature
is compatible (once the two axes are rescaled accordingly)
 with the PDF of the time--filtered logarithm of the curvature (which was studied in \cite{xu}): the reason of 
the convergence of the two procedures may be due to the fact that, by taking the 
logarithm of $\kappa$, the two terms, the one coming from $a_{\perp}$
and the other coming from $u^2$, contribute now linearly to the curvature, 
as $\log \kappa = \log(a_{\perp}) - \log(u^2)$; hence 
when one performs the time average (which is a linear operation) the events at low 
$u^2$ values, characterized by short autocorrelation times, are selectively suppressed, 
similarly to what happens with the introduction of the cutoff. 
It seems then reasonable to infer that such change of the slope
must be attributed to the presence of vortical structures, thus being
connected to small scales  turbulent properties. 
Indeed, because of the constraint imposed on $u^2$, the highest
curvature values are now due to large and temporally persistent
accelerations (or, more correctly, to centripetal accelerations).  If
this is true, the large tail behaviour should be captured by the same
multifractal arguments that have been proved successful for describing
unconditional acceleration statistics \cite{biferale-multi}, which is
highly intermittent \cite{laporta,pinton,biferale-multi}.  We can
write for the curvature PDF (denoted, from now on, with
$\mathcal{P}(\kappa)$, for the sake of simplicity) the following
expression
\begin{equation}
 \mathcal{P}(\kappa) = \int\int 
 \mathcal{P}_{joint}(a,u^2)\delta\left(\kappa-\frac{a}{u^2}\right)
  \Theta(u^2-\Lambda)da du,      
\end{equation}
where $\mathcal{P}_{joint}(a,u^2)$ is the joint PDF of $a$ and
$u^2$,  and the Heaviside $\Theta$ function has been included to take
into account the cutoff on the velocities. 
Note that we use $a$ for $a_\perp$, neglecting geometrical constraints. 
Following the multifractal description presented in  \cite{biferale-multi} 
we may estimate the joint PDF from the relation 
\begin{equation}
 \mathcal{P}_{joint}(a,u^2) = \delta(a-a(h,u))\, \mathcal{P}(u^2); 
 \qquad  a(h,u) \sim \nu^{\frac{2h-1}{1+h}} u^{\frac{3}{1+h}}
 L_0^{-\frac{3h}{1+h}} 
\end{equation}
where $\nu$ is the viscosity, $L_0$ is the integral scale (we will
 keep $L_0 \sim 1$ in the following) and $h$ the
scaling exponent characterising the local velocity fluctuation.
Let us stress that in the expression of $a(h,u)$ a possible over-all stochastic 
prefactor, Reynolds independent, can be included, in order to change the degree of correlation 
between the acceleration and the local velocity field. Such a
 prefactor would  change a bit the joint PDF, 
but not  the global shape of the result.
At the Kolmogorov scale, the probability to have an exponent $h$ is given by \cite{biferale-multi}:
\begin{equation}
\mathcal{P}(h)\sim\parfrac{\tau_\eta}{T_L}^{\frac{3-D(h)}{1-h}}
\end{equation}
 where $T_L$ is the Lagrangian integral time scale and $D(h)$ is the
 fractal dimension of the set characterised by the  exponent $h$. Using for the
 probability density of $u^2$ a third order $\chi^2$ distribution:
\begin{equation}
\mathcal{P}(u^2)\propto u^2e^{-u^2/2}
\end{equation}
we obtain:
\begin{equation} 
\mathcal{P}(\kappa) \sim \int da \int du \int dh
\left(\frac{\nu}{u}\right)^{\frac{3-D(h)}{1+h}} u^2
e^{-\frac{u^2}{2}}
\delta(a-\nu^{\frac{2h-1}{1+h}}u^{\frac{3}{1+h}})\delta\left(\kappa-\frac{a}{u^2}\right)\\
\Theta(u^2-\Lambda).
\end{equation}
Integration over $a$ and $u$ yields the final expression 
\begin{equation} \label{pdfk.mf.analytic}
\mathcal{P}(\kappa)\sim \int_{h_{min}}^{h_{max}}
 \left(\frac{1+h}{1-2h}\right)
\nu^3
\kappa^{\frac{5h-1+D(h)}{1-2h}}exp\left(-\frac{\nu^2\kappa^{\frac{2(1+h)}{1-2h}}}{2}\right)
 dh
\end{equation}
valid for $\kappa\gsim2.0\sigma_\kappa$ (correspondingly $\kappa\eta
  \gsim 0.03$).\footnote{This domain
  limitation arises from the calculation done with the $\Theta$ function.}
Working out the integral with $h\in[0.16,0.38]$ (for integrability
reasons exposed in \cite{biferale-multi}) and using for $D(h)$ the
empirical form  proposed by She and L\'ev$\hat{e}$que \cite{she}
  (which was also shown to fit very well experimental data \cite{xu:2006}, at least 
in the range of singularity exponents $h$ that we are spanning):
\begin{equation}
 D(h)=1+p(h)\left(h-\frac{1}{9}\right)+2\left(\frac{2}{3}\right)^{\frac{p(h)}{3}}
\end{equation}
(where $p(h)=(3/\ln(2/3))\ln[(9h-1)/6\ln(3/2)]$), we have the result
shown in figure (\ref{fig:pdfktail}), in excellent agreement with the
measured PDF for large curvatures. More in detail, it is worth
remarking that the two curves are almost coincident for values of
curvature larger than the ``critical'' value $\tilde{\kappa}$.
Let us notice that the good agreement between data and the
multifractal formalism only after the introduction of a cutoff in the
single point velocity is not an accident. 
We must expect that phenomenological models based on the energy
cascade mechanism work well only for those events that are not
dominated by smooth quasi-laminar fluctuations, 
as it would be the case for the whole curvature statistics without cutoff.

\section{Joint statistics of curvature and torsion}   \label{sec:joint_stat} 
Torsion measurements are very difficult
in experiments \cite{xu} and available only at very low Reynolds
numbers in previous numerical works \cite{braun}. 
The difficulty is due to the fact that they involve the time derivative
of the acceleration. In our numerical data--base acceleration is
differentiable inside mesh grids and continuous everywhere: we could
therefore measure torsion  with a good signal to noise ratio.

\begin{figure}
\begin{center}
\subfigure[]{
\resizebox*{7cm}{!}{\includegraphics{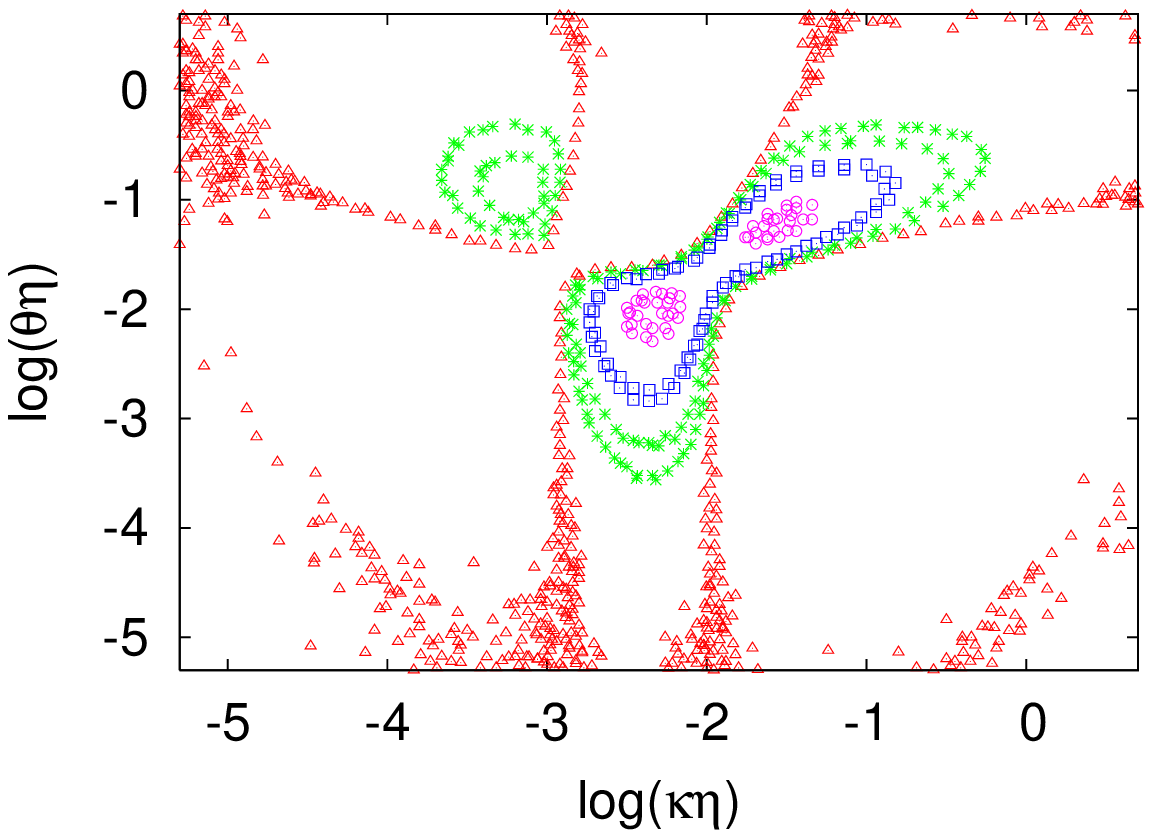}}}%
\subfigure[]{
\resizebox*{7cm}{!}{\includegraphics{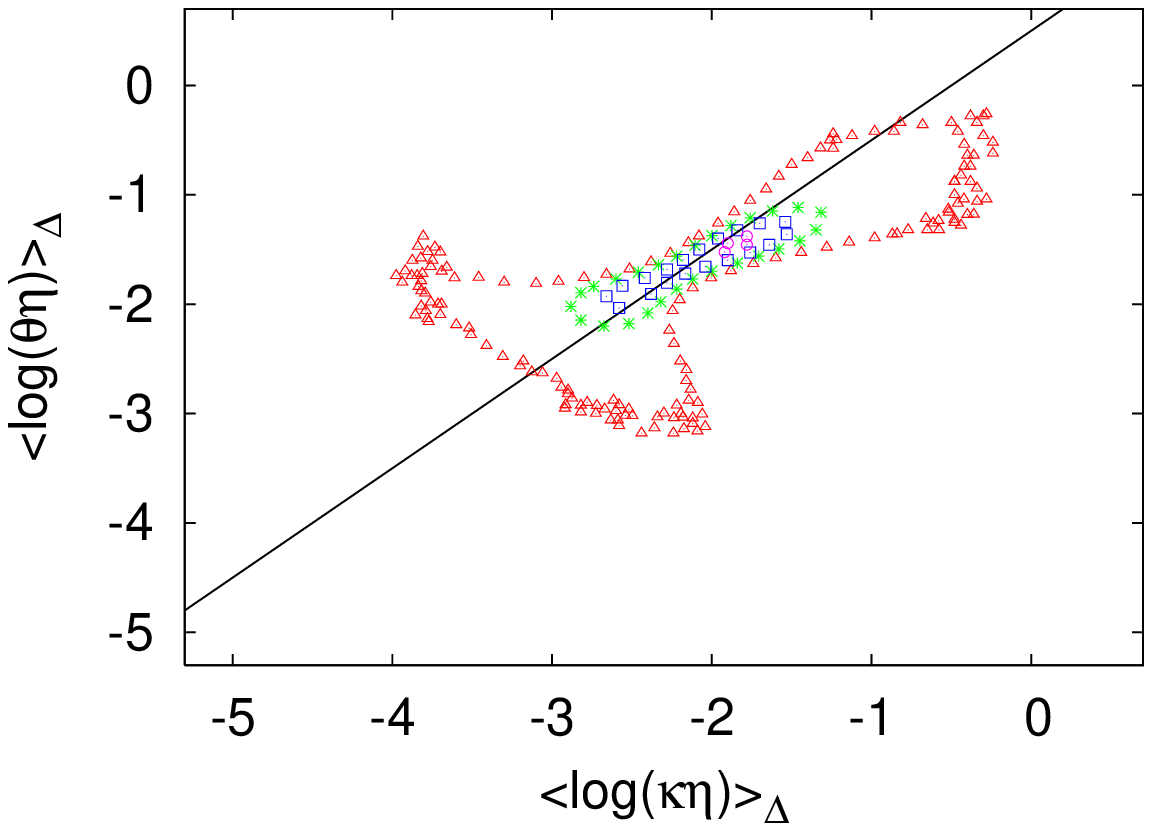}}}%
\caption{\label{}\textit{Residuals} of the logarithms of curvature and
 torsion, instantaneous (a) and time--averaged (b) 
 over a time window $\Delta=10\tau_\eta$,
 plotted as a 2D map with isocontours (only $\mathcal{R}>0$ is
 shown). Correlation is lesser on exterior contours. In the right
 panel we also plot a straight line  parallel to the
plane bisector $\log(\theta)=\log(\kappa)$.}%
\label{fig:remn}
\end{center} 
\end{figure}

\begin{figure}
\begin{center}
\subfigure[]{
\resizebox*{7cm}{!}{\includegraphics{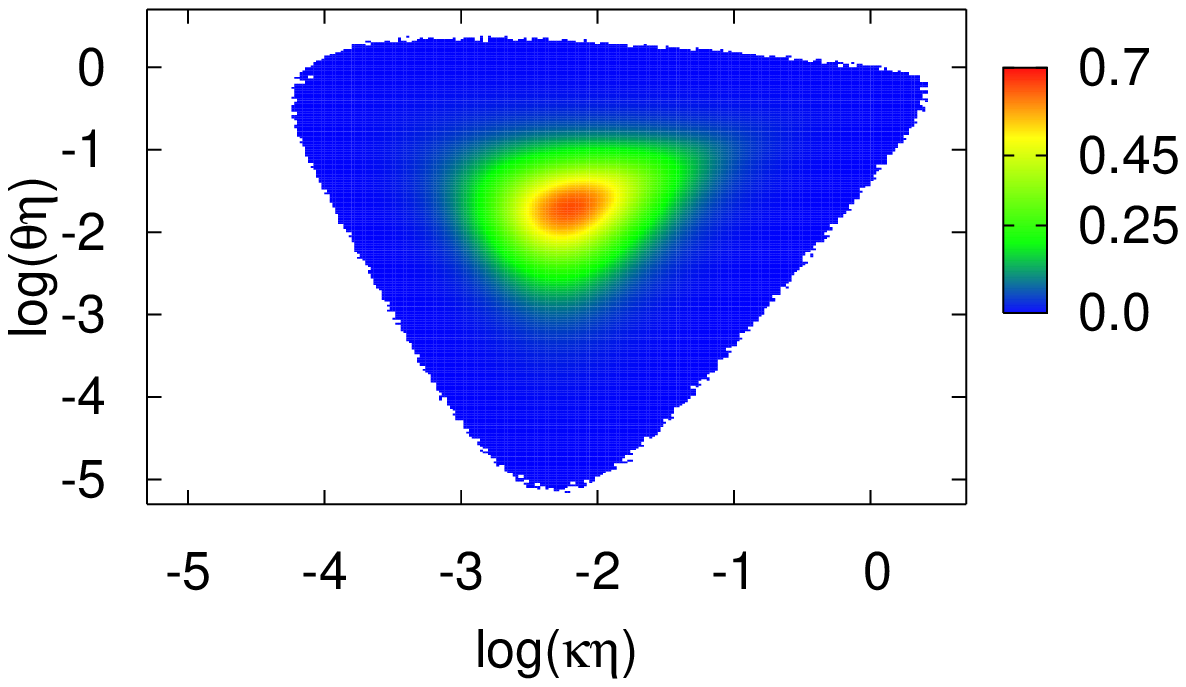}}}%
\subfigure[]{
\resizebox*{7cm}{!}{\includegraphics{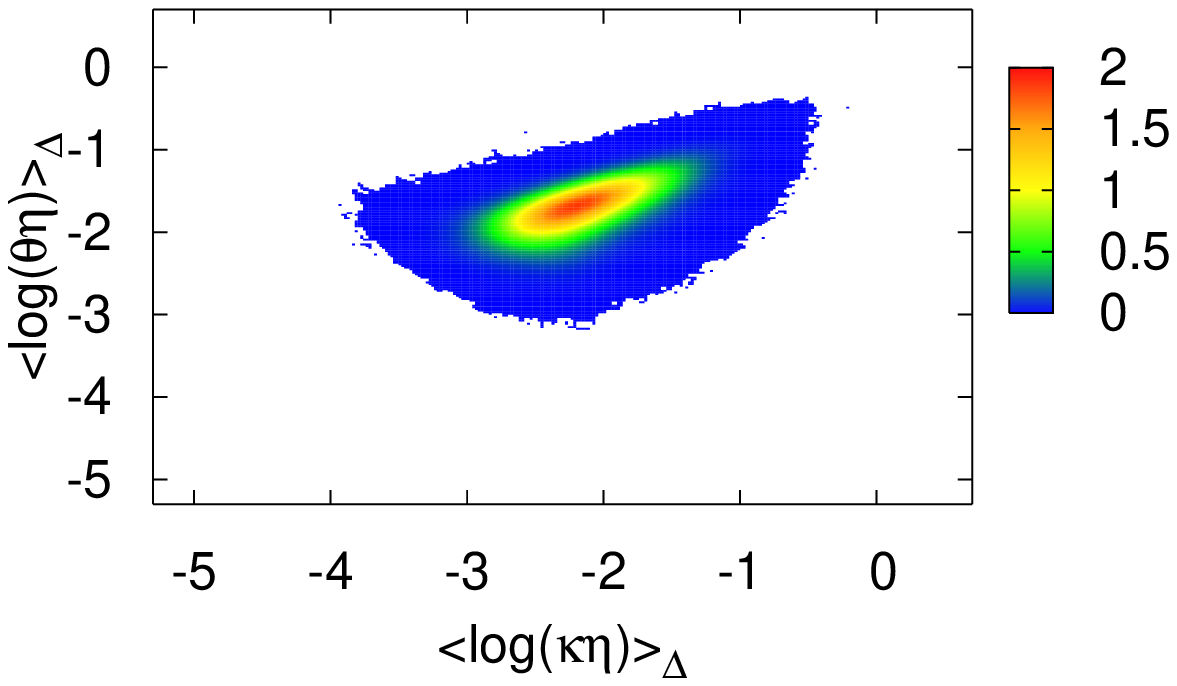}}}%
\caption{\label{}Joint probability distribution functions of the
  logarithms of curvature and torsion,
  instantaneous (a) and time--averaged (b) over a time window $\Delta
  = 10\tau_\eta$.}%
\label{fig:jpdf}
\end{center} 
\end{figure}

It is natural to suppose that the emergence of coherent structures with
peculiar geometries induces somehow a correlation between curvature
and torsion of trajectories, as they contain informations on such
structures.    
We studied, then, the joint
statistics of the two quantities. We calculated the
\emph{residual value} \cite{bif-toschi}  of the logarithms of $\kappa$ and $\theta$, that
is the quantity
\begin{equation}
 \mathcal{R}(\langle\log(\kappa)\rangle_\Delta,\langle\log(\theta)\rangle_\Delta)\eqdef
 \mathcal{P}_{joint}(\langle\log(\kappa)\rangle_\Delta,\langle\log(\theta)\rangle_\Delta)
 -\mathcal{P}(\langle\log(\kappa)\rangle_\Delta)\mathcal{P}(\langle\log(\theta)\rangle_\Delta)
\end{equation}
which quantifies the degree of correlation existing between the two
variables. In figure (\ref{fig:remn}) we plot
$\mathcal{R}(\log(\kappa),\log(\theta))$ 
and $\mathcal{R}(\langle\log(\kappa)\rangle_\Delta,\langle\log(\theta)\rangle_\Delta)$,
with $\Delta=10\tau_\eta$, projected onto the
$\log(\theta)$ \textit{vs} $\log(\kappa)$ plane (isocontours are drawn
for positive correlation $\mathcal{R}>0$).
In figure (\ref{fig:jpdf}) we also show the original joint PDF 
$\mathcal{P}_{joint}(\langle\log(\kappa)\rangle_\Delta,\langle\log(\theta)\rangle_\Delta)$
for both the instantaneous and time--filtered quantities.
The most relevant feature to be stressed is that the effect of time
filtering is to accumulate the isocontours  to
intensify the correlation along the straight lines
$\log(\theta)=\log(\kappa)+c$ (parallel to the plane bisector). This
equation is exactly what one obtains by taking the logarithm of
\begin{equation}
\theta=\frac{\lambda}{R}\kappa
\end{equation}
which is the relation linking torsion and curvature for a
``perfect'' cylindrical helix ($\lambda$ being its reduced step and
$R$ the cylinder radius), that is the prototype of spatial curve that
we imagine when thinking of a vortex tube. 
From the value of the intercept we can then also estimate the ratio
between the step and the radius of the ``mean'' helix to be $\lambda/r
\sim 3 $. These two geometrical parameters are related to the inverse
curvature (i.e. the radius of curvature) via
\begin{equation}\label{eq:roc}
R=\frac{1}{\kappa}=\frac{r^2+\lambda^2}{r};
\end{equation}
furthermore from the critical value at which the change of slope of the
curvature PDF occurs (discussed in the previous section), we get for
the radius of curvature $R\sim 10 \eta$. Hence, substituting the
measured ratio $\lambda/r$ into (\ref{eq:roc}), we obtain, as an estimate
for the radius of the helix,
\begin{equation}
r \sim \eta.
\end{equation}
These observations strengthen
furthermore the conjecture that when filtering and avoiding the
very low velocity events, the study of the statistics of curvature and
torsion provides an insight on the signatures of vortex filaments.
 
\section{Conclusions}
We have analyzed data from DNS of Lagrangian tracers, focusing our
attention on the statistics of parameters able to characterize the
geometry of particle trajectories, namely curvature and torsion. We
found that the statistics of instantaneous curvature is mainly
dominated by large scale flow reversal events, rather than by small
scale structures like vortex tubes as also found in experimental works in \cite{xu}. 
Indeed, the asymptotics of the PDF
can be reproduced by simple Gaussian statistics arguments \cite{xu}.
In order to unravel small-scales turbulence features, we meant to
compute the PDF of curvature averaged in time, including only those
events for which the squared velocity magnitude $u^2$ was greater than
a certain threshold, so to avoid the reversals (where $u^2\sim0$) and
to exalt the statistical signature of vortices. This approach
successfully resulted in a marked change of slope in the PDF tail at
high curvature values, pointing out the possible presence of vortex
filaments.  The criterion here presented is therefore different from
what proposed in other works \cite{moisy}. Moreover, we could derive
an analytical expression for $\mathcal{P}_\Lambda(\kappa_{\Delta})$,
in the context of the multifractal formalism, which fits well the far
tails, providing an interesting matching between geometry and
phenomenology. The analysis of joint statistics of curvature and
torsion has been also addressed, confirming our statements on the
topology of coherent small scale structures.

\section*{Acknowledgements} 
I gratefully thank Luca Biferale for his precious suggestions and
support. I would also like to acknowledge
useful and interesting discussions with Federico Toschi and Alessandra
Lanotte.

\clearpage


\end{document}